\documentclass[epj,twocolumn]{webofc}
\usepackage[varg]{txfonts}   % Web of Conferences font
\usepackage{color}
\pdfoutput=1
\woctitle{Dark Matter, Hadron Physics and Fusion Physics}
\newcommand {\pT}{\ensuremath{p_{\mathrm{T}}}}
\newcommand {\meanpT}{$\langle \pT \rangle$}

\newcommand {\Npart}{$\langle N_{\mathrm{part}} \rangle$}

\newcommand {\modrap} {$\left | y \right | $}
\newcommand {\dndy}{d$N$/d$y$}

\newcommand {\dnchdeta}{d$N_{\mathrm{ch}}$/d$\eta$}
\newcommand {\dedx}{d$E$/d$x$}
\newcommand {\PbPb}{\mbox{Pb--Pb}}
\newcommand {\pPb}{\mbox{p--Pb}}
\newcommand {\pp}{pp}
\newcommand {\s}{$\sqrt{s}$}
\newcommand {\sNN}{$\sqrt{s_{\mathrm{NN}}}$}

\newcommand {\RAA}{$R_{\mathrm{AA}}$}
\newcommand {\RpPb}{$R_{\mathrm{pPb}}$}
\newcommand {\gmom} {GeV/$c$}

\newcommand {\fmc} {fm/$c$}
\newcommand{\K}{\mbox{$\mathrm {K}$}}
\newcommand{\p}{\mbox{$\mathrm {p}$}}
\newcommand{\ap}{\mbox{$\mathrm {\overline{p}}$}}
\newcommand{\pion}{\mbox {$\mathrm {\pi}$}}
\newcommand{\Kzs}{\mbox{$\mathrm {K^0_S}$}}
\newcommand{\rmLambda}{\mbox{$\mathrm {\Lambda}$}}

\newcommand{\rmXi}{\mbox{$\mathrm {\Xi^{-}}$}}

\newcommand{\rmOmega}{\mbox{$\mathrm {\Omega^{-}}$}}

\newcommand{\kstar}{\mbox{\K$^{*}$(892)$^{\mathrm{0}}$}}
\newcommand{\akstar}{\mbox{$\overline{\K^{*}}$(892)$^{\mathrm{0}}$}}
\newcommand{\phir}{\mbox{$\mathrm {\phi}$(1020)}}
\newcommand{\rmphi}{\mbox{$\mathrm{\phi}$}}
\newcommand{\simplekstar}{\mbox{\K$^{*0}$}}

\begin{document}
\selectlanguage{english}
\title{Recent ALICE results on hadronic resonance production}

\author{A.~Badal\`a~\inst{1}\fnsep\thanks{\email{Angela.Badala@ct.infn.it}} for the ALICE Collaboration
        % etc.
}

\institute{INFN - Sezione di Catania , Via S. Sofia 64, 95123, Catania (Italy) 
          }

\abstract{%
Hadronic resonances are a valuable tool to study the properties of the medium formed in heavy-ion collisions.  In particular, they can provide information on particle-formation mechanisms and on the properties of the medium at chemical freeze-out. Furthermore they contribute to the systematic study of parton
energy loss and quark recombination. 
Measurements of resonances in pp  and in \pPb~collisions provide a necessary baseline for heavy-ion data and help to disentangle initial-state effects from 
medium-induced effects. In this paper  the latest  ALICE results on mid-rapidity  \kstar~and \phir~production in \pp, \pPb~and  \PbPb~collisions at  LHC energies are presented. 
}
\maketitle
\section{Introduction}
\label{intro}
\begin{figure*}
\centering
%\vskip -0.65cm
\includegraphics[width=8.0cm]{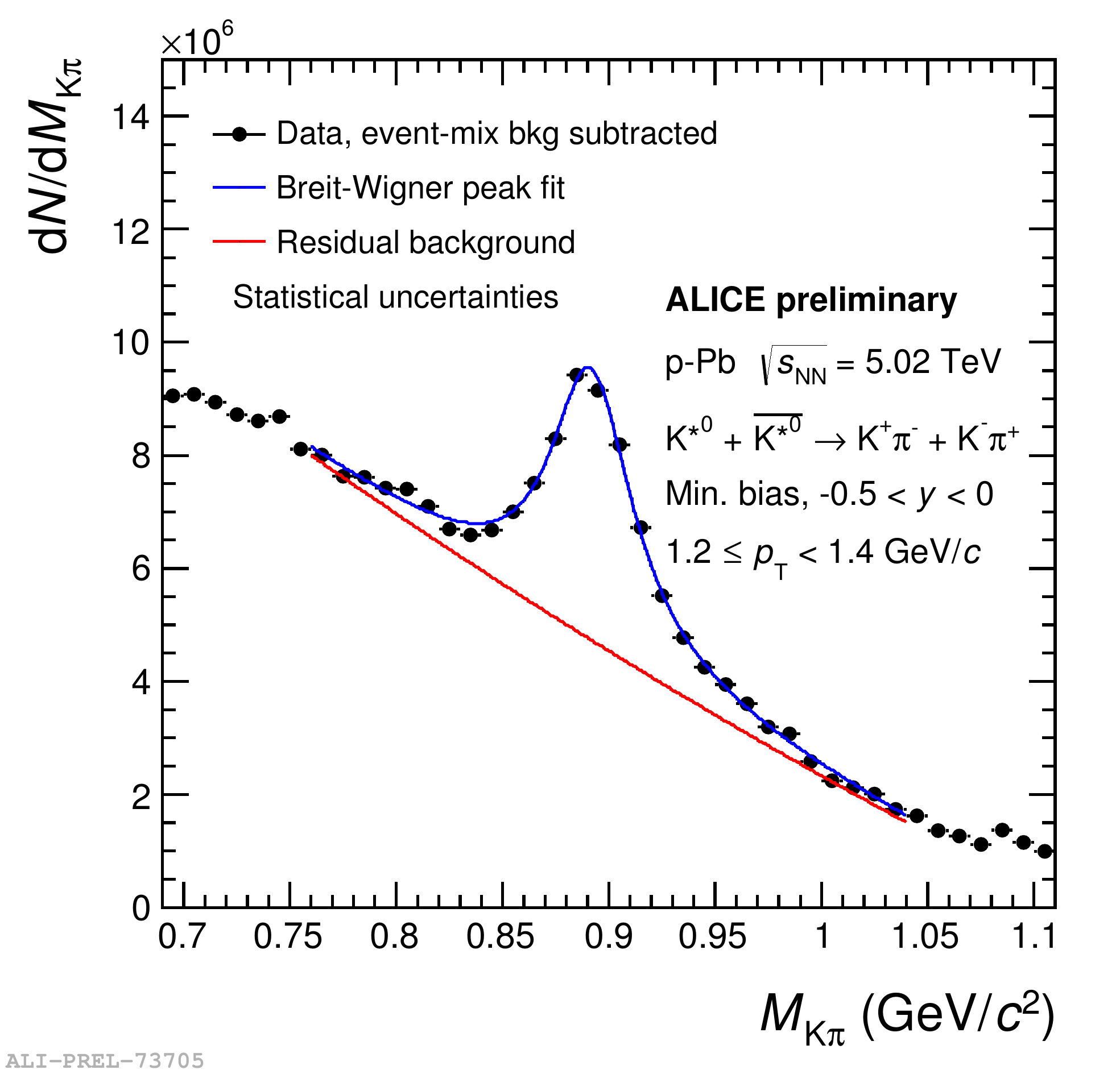}
\includegraphics[width=8.0cm]{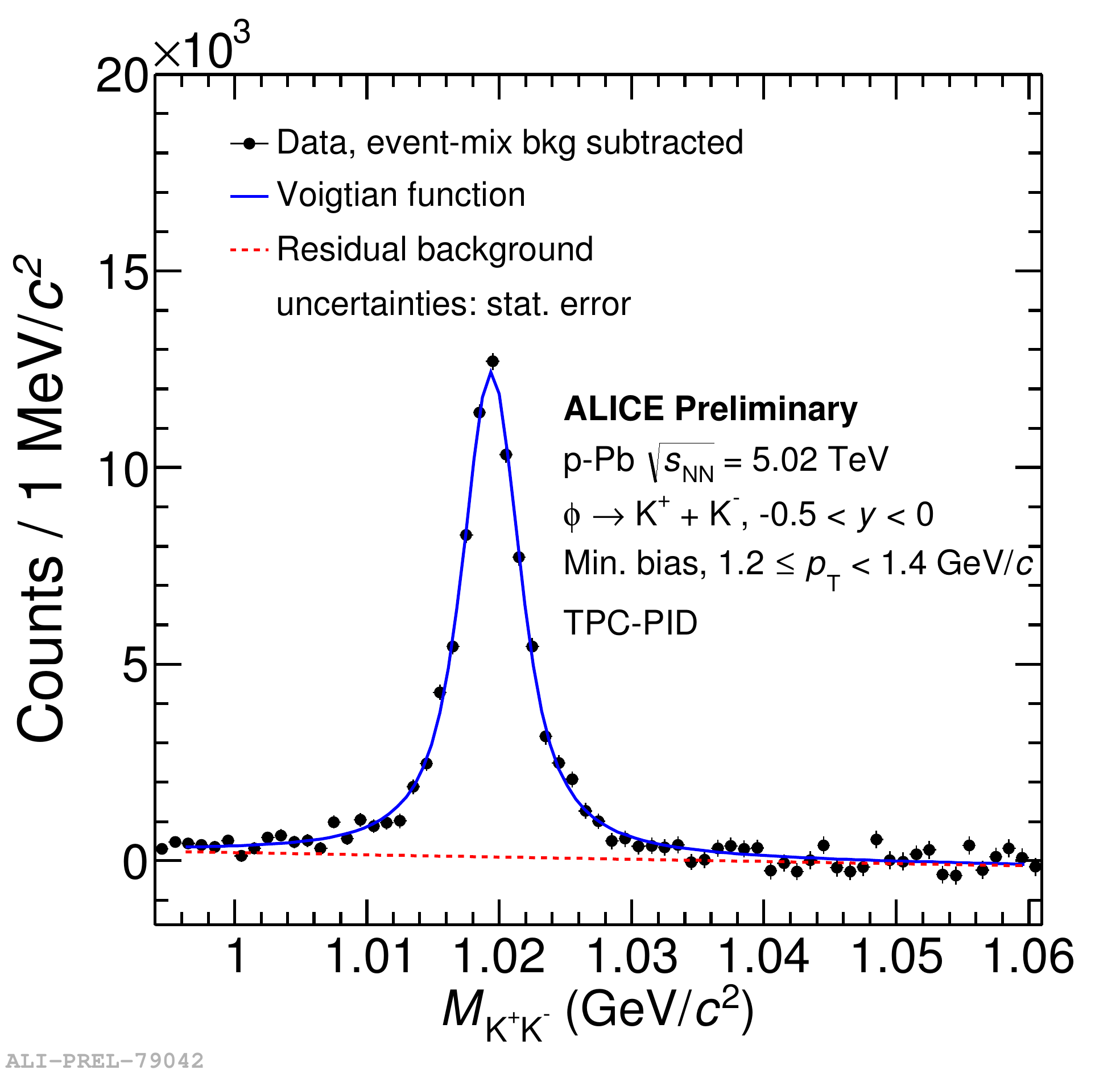}
\caption{ (Left panel)The \K\pion~invariant mass distribution in \pPb~collisions at \sNN~=~5.02~TeV in the \pT~range 1.2~$<$~\pT~$<$~1.4~\gmom, after the subtraction of the combinatorial background.  The solid curve represents the fitting function, the dashed curve describes the residual background.  The fitting function is the sum of a relativistic \p-wave Breit-Wigner function with a Boltzamann factor to account for phase space~\cite{Breit} and a polynomial.
(Right panel) The \K$^+$\K$^-$ invariant mass distribution in \pPb~collisions at \sNN~=~5.02~TeV in the \pT~range 1.2~$<$~\pT~$<$~1.4~\gmom, after the subtraction of the combinatorial background.  The solid curve represents the fitting distribution, the dashed curve describes the residual background.  The fitting function is the sum of a Voigtian function and a polynomial.}
\label{fig:invmass}       % Give a unique label
\end{figure*}

\begin{figure*}
\centering
%\begin{figure}[h]
%\begin{minipage}{32pc}
%\vskip -0.1cm
\includegraphics[width=32pc]{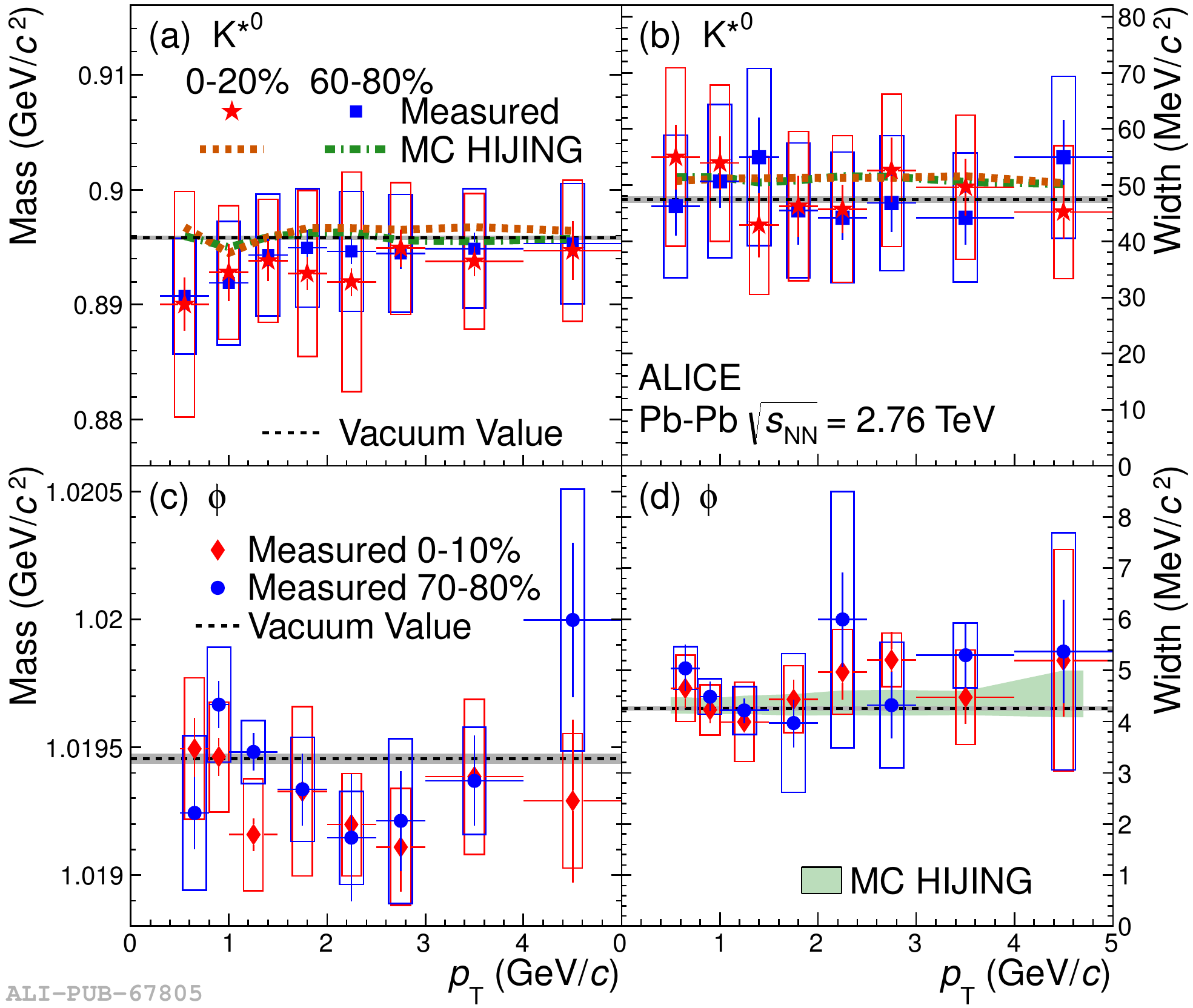}
\caption{\label{fig:mass_width} Measured \simplekstar~mass (a) and width (b) in \PbPb~collisions at \sNN~=~2.76~TeV in the 0-20\% and 60-80\% centrality intervals. Measured \rmphi~mass (c) and width (d) in \PbPb~collisions at \sNN~=~2.76~TeV  in the 0-10\% and 70-80\% centrality intervals.  Statistical uncertainties are shown as bar and systematic uncertainties are shown as boxes. Mass (for \simplekstar) and width (for \simplekstar~and \rmphi) extracted from Monte-Carlo HIJING simulations are also shown. The vacuum value of the \simplekstar~and \rmphi~mass and width~\cite{pdg} are indicated by the horizontal dashed lines.}
\end{figure*}
%\end{minipage} 
%\end{figure}
Hadronic resonances are sensitive probes for the different phases of the evolution of the medium produced in ultrarelativistic heavy-ion collisions. Their lifetimes are, in fact, in the range of few \fmc~to some tens of \fmc, which is comparable with the one  ($\sim$10 \fmc) estimated for the created fireball. 
In these collisions a hot and dense state of matter~\cite{qgp1,qgp1n,qgp2}, the quark-gluon plasma, is expected to be produced. At a critical temperature of about 160~MeV a cross-over transition  between the partonic (i.e. a system with deconfined quarks) and hadronic phases is expected to take place.   After the chemical freeze-out the system continues to expand and to cool until the kinetic freeze-out is reached.  Modifications of the yield, the mean transverse momentum \meanpT~and ratio of the yields of resonances to stable particles  can provide information about the regeneration and re-scattering effects in the hadronic phase. In fact the final recostructible resonance yields depend not only on the chemical freeze-out temperature but also on the scattering cross section of the resonance decay products and the  timescale between the chemical and the kinetic freeze-out, which control the fraction of 'undetected' particles. However resonances may be regenerated by pseudo-elastic interactions in the hadronic medium, a process driven by the cross-section of the interacting hadrons.

Moreover, due to the expected (partial) restoration of the chiral symmetry in the formed quark-gluon-plasma, modifications of the properties of the resonances (mass, width and branching ratio) have been predicted~\cite{chiral1,chiral2}.

In heavy ion collisions information on particle formation mechanisms can be derived from the comparison 
of particles with similar mass but different baryon number and/or strangeness content. Comparison of resonance production with that of long lived hadrons can be interesting in this respect. In particular it is worth studying the \rmphi~meson with its hidden strangeness content. Finally, a contribution to the systematic study of the in-medium parton energy loss can be obtained from the measurement of the resonance production at high-\pT.

Both meson and baryon resonances have been measured by the ALICE experiment~\cite{alice} in different collisions systems (pp, \pPb, \PbPb) 
at LHC energies~\cite{phi900,phi_kstar_pp,phi_kstar_pbpb,Sigmastar_pp}. Resonance measurements in \pp~and \pPb~systems are useful as references and to disentangle initial-state effects from genuine in-medium effects, which may occur in \PbPb~collisions. In this paper, focus is given to the meson resonances \kstar~and \phir, reconstructed at mid-rapidity in \PbPb~collisions at  \sNN~=~2.76~TeV~\cite{phi_kstar_pbpb} and in \pPb~collisions at \sNN~=~5.02~TeV.

\section{Data analysis and resonance reconstruction}
\label{data}
The results reported here  refer to analyses carried out on samples of minimum-bias \pp~data at \s~=~2.76~TeV and  7~TeV (about 33 and 80 million events, respectively)  and of minimum bias \PbPb~data at \sNN~=~2.76~TeV (about 13 million events) and \pPb~events at \sNN~=~5.02~TeV (about 90 million events), collected using the ALICE detector~\cite{alice}. More information about the performance of ALICE can be found in~\cite{alice_perf}.

The \kstar~and \phir~production has been measured by reconstructing the resonances through their main hadronic decay: \kstar $\longrightarrow $ \K$^+\pi^-$ 
(and corresponding antiparticle) and \rmphi$\longrightarrow$\K$^+$\K$^-$. 
All measurements of \kstar~and \akstar~are averaged and these mesons are referred to as \simplekstar~in the following.
In \pp~and in \PbPb~collisions resonances are measured in one unit of rapidity 
 \modrap~$<$~0.5 in the centre-of-mass reference frame, while in \pPb~the  rapidity range is restricted to 
 -0.5~$<y<$~0, in order to ensure the best detector acceptance with the shifted centre-of-mass of the system.
 The position of the primary vertex  is estimated using the tracks reconstructed in the Inner Tracking System (ITS)  and in the Time Projection Chamber (TPC) and 
 its component along the beam axis is required to be within 10~cm from the centre of the ALICE detector. The VZERO detectors, two scintillator hodoscopes covering the pseudo-rapidity 
 ranges 2.8~$<\eta<$~5.1 (VZERO-A) and -3.7~$<\eta<$~-1.7~(VZERO-C), were used for event triggering and the definition of centrality and multiplicity classes  respectively in \PbPb~\cite{centrality} and \pPb~\cite{multpPb} collisions.
%{\color{red}
Identification of pions and kaons is carried out using the measurement of the specific energy loss \dedx~in the TPC. The TPC \dedx~measurements allow pions to be separated from kaons for momenta up to $p \sim$~0.7~\gmom, while the proton/antiproton band starts to overlap with the pion/kaon band at $p \sim$~1~\gmom.  An improvement in the significance of the signal has been achieved using the information from the Time-Of-Flight (TOF) detector, for tracks for which it is available. The TOF allows pions and kaons to be unambiguously identified up to $p \sim$~1.5-2.0~\gmom.  The two mesons can be distinguished from (anti)protons up to $p \sim$~2.5~\gmom.
%} 

Resonances were reconstructed by computing the invariant mass spectrum of all primary track pairs and then subtracting the combinatorial background, estimated by event-mixing or like-sign techniques. The signal of the invariant mass distribution of \simplekstar(\rmphi) was fitted with a relativistic Breit-Wigner (Voigtian) function~\cite{Breit} plus a polynomial for the residual background.
In Fig.~\ref{fig:invmass} some examples of invariant mass spectra for \simplekstar~and \rmphi~after the subtraction of the
residual background in \pPb~collisions at \sNN~=~5.02~TeV are shown.
Examples of invariant mass spectra in \pp~and \PbPb~collisions can be found in~\cite{phi_kstar_pp,phi_kstar_pbpb}.
In all the collision systems the mass and width of \simplekstar(\rmphi) are found to be close to the vacuum value~\cite{pdg}. In particular, as can be seen in Fig.~\ref{fig:mass_width}, within the errors, no mass shift or broadening has been observed in \PbPb~collisions .

\section{Results} 
\label{results}
The procedure used to estimate the \simplekstar~and \rmphi~yield has been extensively explained  in~\cite{phi_kstar_pp,phi_kstar_pbpb}.  In \pPb~collisions to extract the particle yields and the \meanpT,  the spectra are fitted using a  L\'evy-Tsallis parameterization~\cite{Tsallis}. To extract the \dndy~the measured \pT~distributions are integrated, while the fits are used to estimate the resonance yield at low and high \pT,  where no signal could be measured.   It may be noted that the extrapolated fraction of the total yield for the \simplekstar~is lower than 0.1$\%$. 

\subsection{Mean transverse momentum}
\label{meanpt}

\begin{figure}
\centering
\includegraphics[width=8cm]{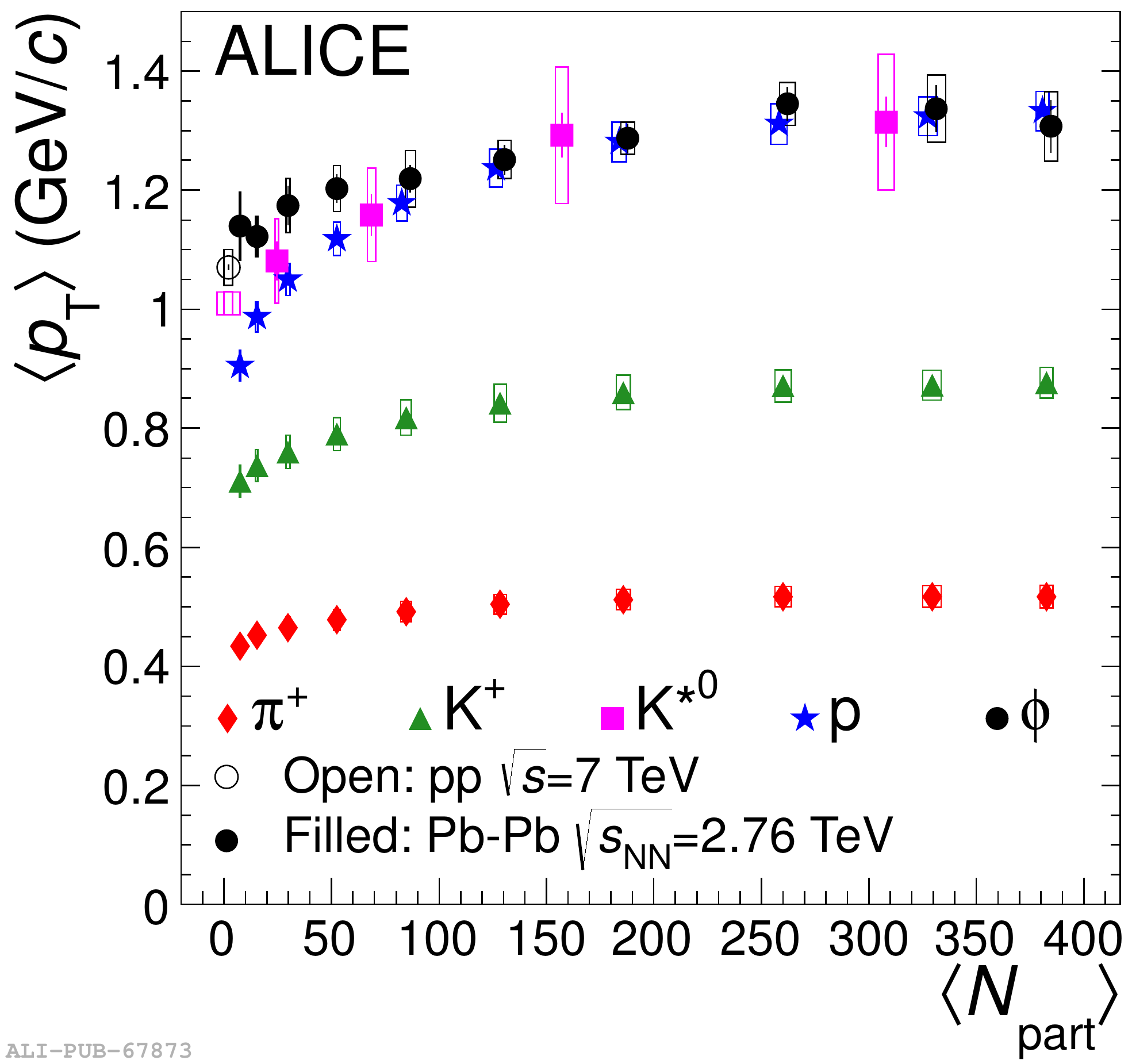}
\caption{Mean transverse momentum of \pion$^+$, \K$^+$, \simplekstar, \p~and \rmphi~in \PbPb~collisions at \sNN~=~2.76~TeV (filled symbols)~\cite{blast_ALICE} as a function of mean number of participant \Npart.  The open symbols represent the \meanpT~values for the resonances in pp collisions at \s~=~7~TeV~\cite{phi_kstar_pp}. }
\label{fig:meanpT_PbPb}       % Give a unique label
\end{figure}

Information on the particle production mechanisms can be obtained from the mean transverse momentum, \meanpT. This quantity has been measured for various particles 
(\pion$^+$, \K$^+$, \Kzs, \p, \rmLambda, \rmXi, \rmOmega~and their anti-particles)~\cite{multpPb} and for \simplekstar, \rmphi~in \pp, \pPb~and \PbPb~collisions~\cite{phi_kstar_pp,phi_kstar_pbpb,Bellini_QM2014,blast_ALICE,multpPb}.  
In central \PbPb~collisions,  particles with similar mass (\simplekstar, \p~and \rmphi) have
similar \meanpT~(Fig.~\ref{fig:meanpT_PbPb}). This is consistent with hydrodynamical medium evolution, where the \pT~distribution is mainly determined by particle mass.

\begin{figure}
\centering
\includegraphics[width=8cm]{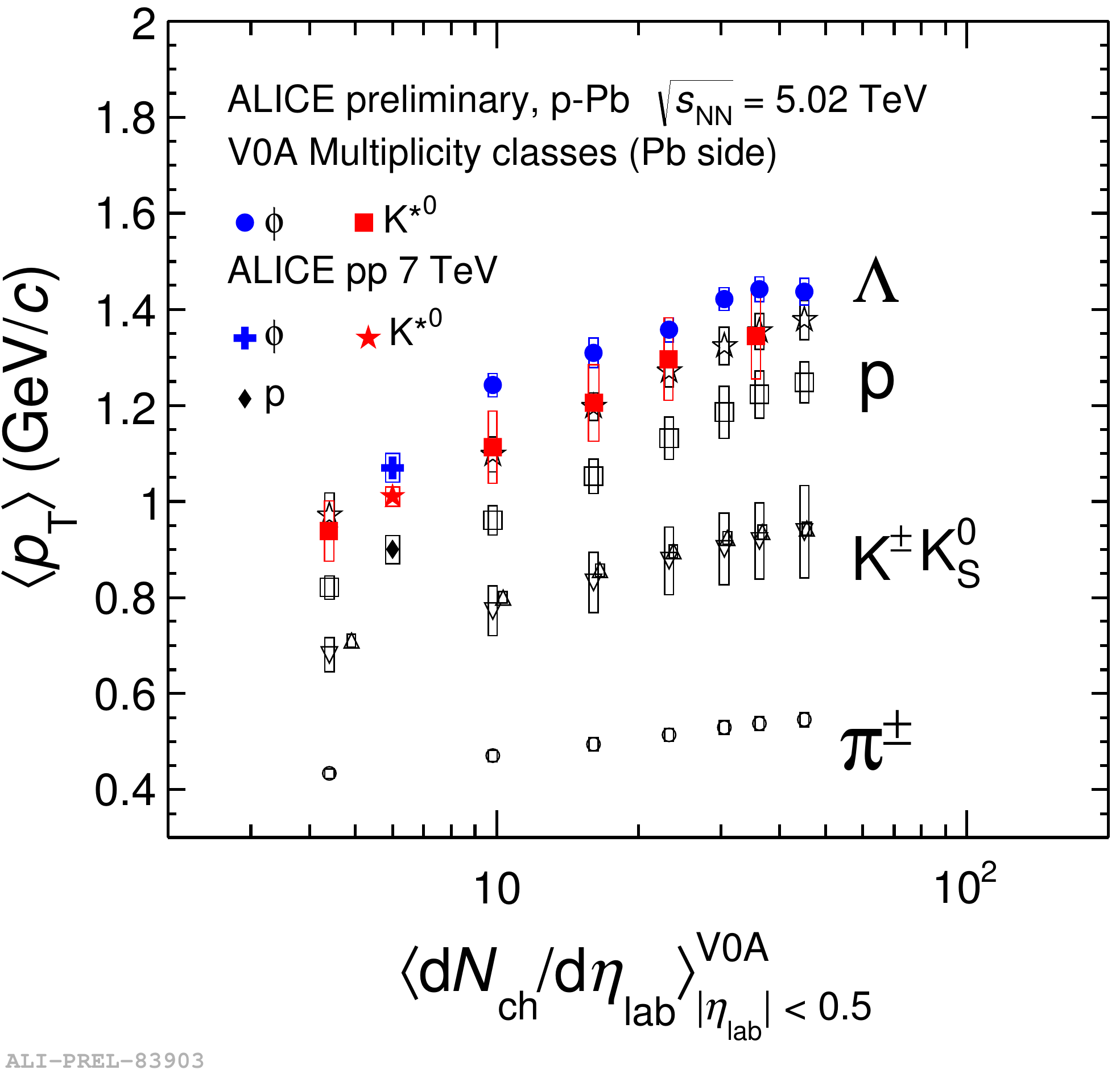}
\caption{Mean transverse momentum of  \pion$^{\pm}$, \K$^{\pm}$, \Kzs, \simplekstar, \p, \rmphi~and \rmLambda~as a function of the  charged particle multiplicity density \dnchdeta~for \pp~and
\pPb~collisions, respectively,  at \s~=~7~TeV and \sNN~=~5.02~TeV.  Filled symbols refer to resonance \meanpT~in \pPb~collisions.}
\label{fig:meanpT_pPb}       % Give a unique label
\end{figure}

\begin{figure}
\centering
\includegraphics[width=8.2cm]{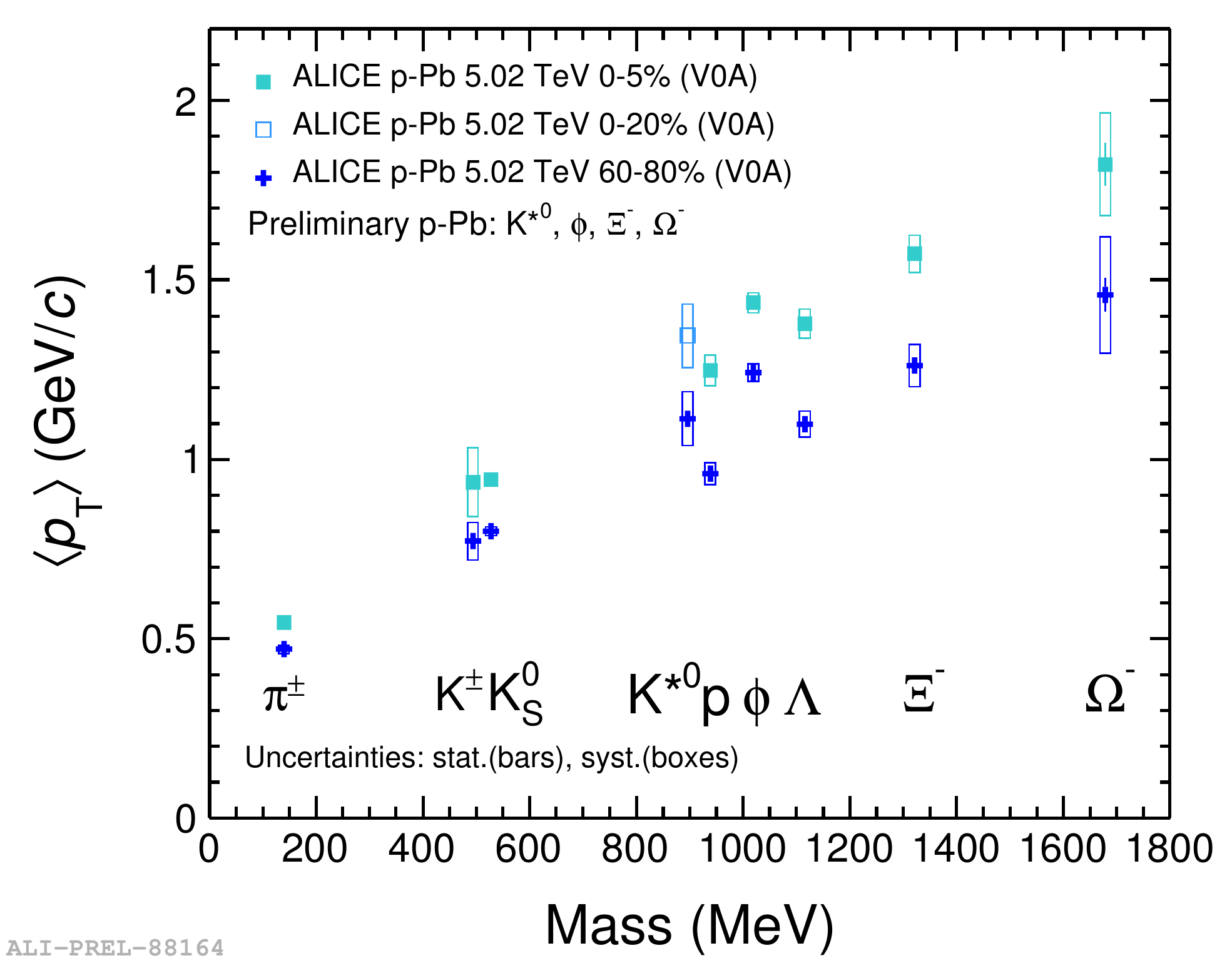}
\caption{Mean transverse momentum as a function of the mass of the detected particles for the 0-5\% and 60-80\% VZERO-A (V0A) multiplicity classes in \pPb~collisions at \sNN~=~5.02~TeV. The V0A multiplicity classes  for \simplekstar~are 0-20\%.}
\label{fig:meanpT_pPb_mass}       % Give a unique label
\end{figure}

\begin{figure*}
\centering
%\begin{figure}[h]
%\begin{minipage}{32pc}
%\vskip -0.1cm
\includegraphics[width=36pc]{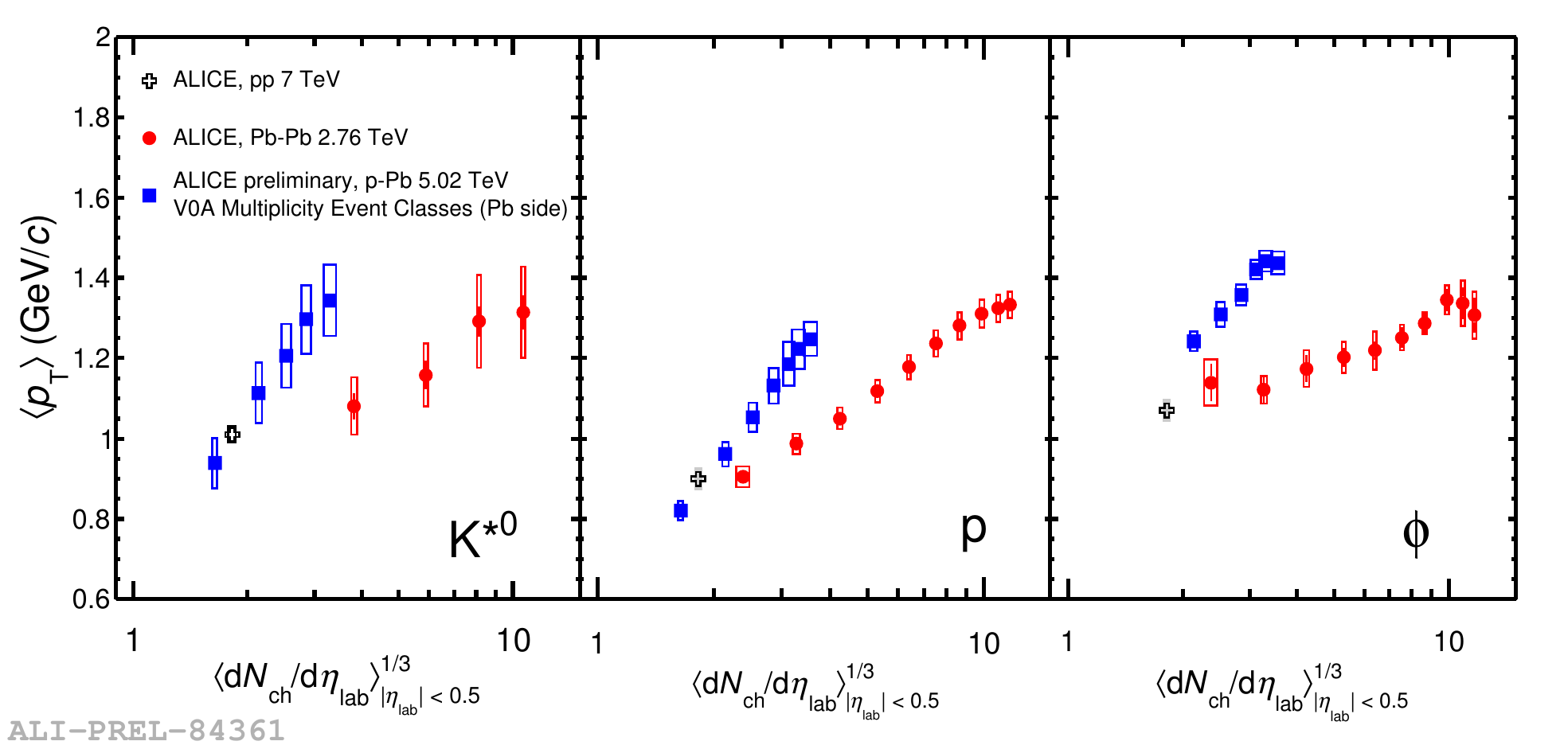}
\caption{\label{fig:meanpT_pPb_mult} System size dependence,  defined as the cubic root of the average charged particle multiplicity density, of the mean transverse momentum of \simplekstar, \p~and \rmphi~in \pp~at \s~=~7~TeV (black crosses), \pPb~at \sNN~=~5.02~TeV (blue squares) and \PbPb~at \sNN~=~2.76~TeV (red squares).}
\end{figure*}

In \pPb~collisions the \meanpT~of resonances increases as a function of the average charged particle multiplicity density \dnchdeta~as for other hadrons (Fig.~\ref{fig:meanpT_pPb}). Moreover while \meanpT~of long lived hadrons follow mass ordering (\meanpT$_{\rmLambda}>$~\meanpT$_{\p}>$~\meanpT$_{\Kzs,\K^{\pm}}>$~\meanpT$_{{\pion}^{\pm}}$), the \meanpT~of \simplekstar~and \rmphi~is larger than that of protons and \meanpT~of \rmphi~is also larger of the \meanpT~of \rmLambda.  A similar trend is observed also in \pp~at 7~TeV~(Fig.~\ref{fig:meanpT_pPb}), where 
\meanpT$_{\rmphi}~>$~\meanpT$_{\simplekstar}~>$~\meanpT$_{\p}$. 
The question remains open whether the mesonic resonances deviate from mass ordering or the baryons, namely \p~and \rmLambda, do, instead.  However Fig.~\ref{fig:meanpT_pPb_mass}, which shows the \meanpT~as a function of the particle mass, including also \rmXi~and \rmOmega~hyperon results,  suggests the possibility of two different trends: one for the mesons (including the resonances) and another for the baryons. 
In Fig.~\ref{fig:meanpT_pPb_mult} the \meanpT~of \simplekstar~and \rmphi~measured in \pp~at \s~=~7~TeV and in \pPb~at 
 \sNN~=~5.02~TeV and in \PbPb~at \sNN~=~2.76~TeV is shown as a function of the cubic root of the average charged particle multiplicity density $\langle$\dnchdeta $\rangle^{1/3}$, a quantity that it  is related to the system size~\cite{radius1,radius2}. 
For all these particles the values of \meanpT~for the highest-multiplicity event class in \pPb~collisions reach (or even exceed) the values measured in central \PbPb~collisions. 
%{\color{red}
The shape of the spectra and the  mass ordering of the \meanpT~in \PbPb~collisions are explained as a consequence of collective hydrodynamical flow.  The observed increase of the \meanpT~as a function of the charged-particle multiplicity observed in \pp~collisions~\cite{meanpT} has been attributed to color reconnection between strings formed by multiple parton-parton interactions.  A similar effect could be present in \pPb~collisions and may explain the deviation of the  observed \meanpT~from the one expected from a  sum of incoherent nucleon-nucleon collisions~\cite{meanpT}. 
Initial and/or final state effects are the responsible of the differences observed between \pPb~and \PbPb~production and need to be further studied.
%}
%All these observations suggests the possibility of  different production mechanisms in \PbPb~and \pPb~collisions~\cite{meanpT}. 

%\end{minipage} 
%\end{figure}
\subsection{Particle ratios}
\label{ratio}
\begin{figure*}
\centering
\vskip -0.65cm
\includegraphics[width=8.5cm]{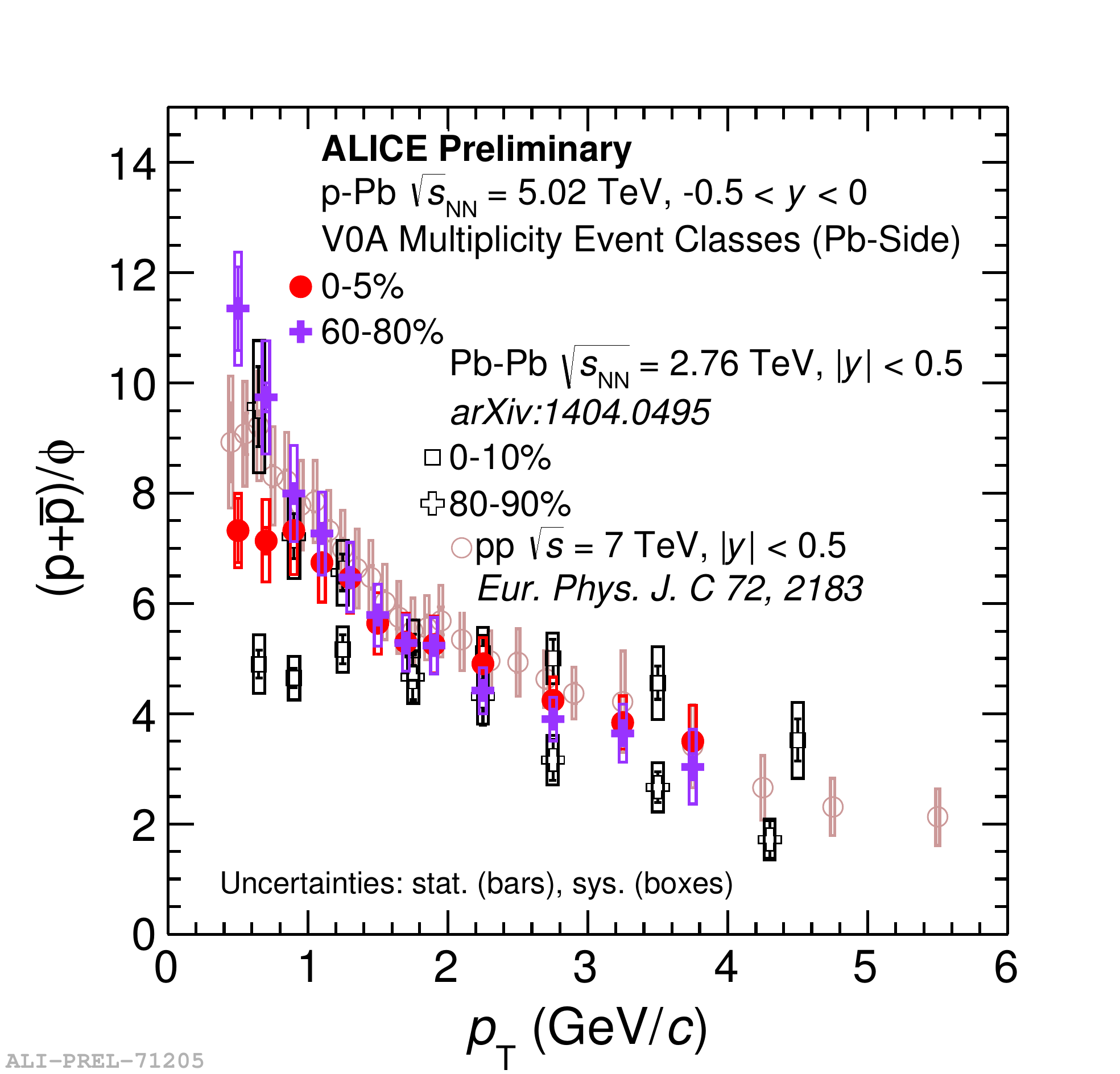}
\includegraphics[width=7.8cm]{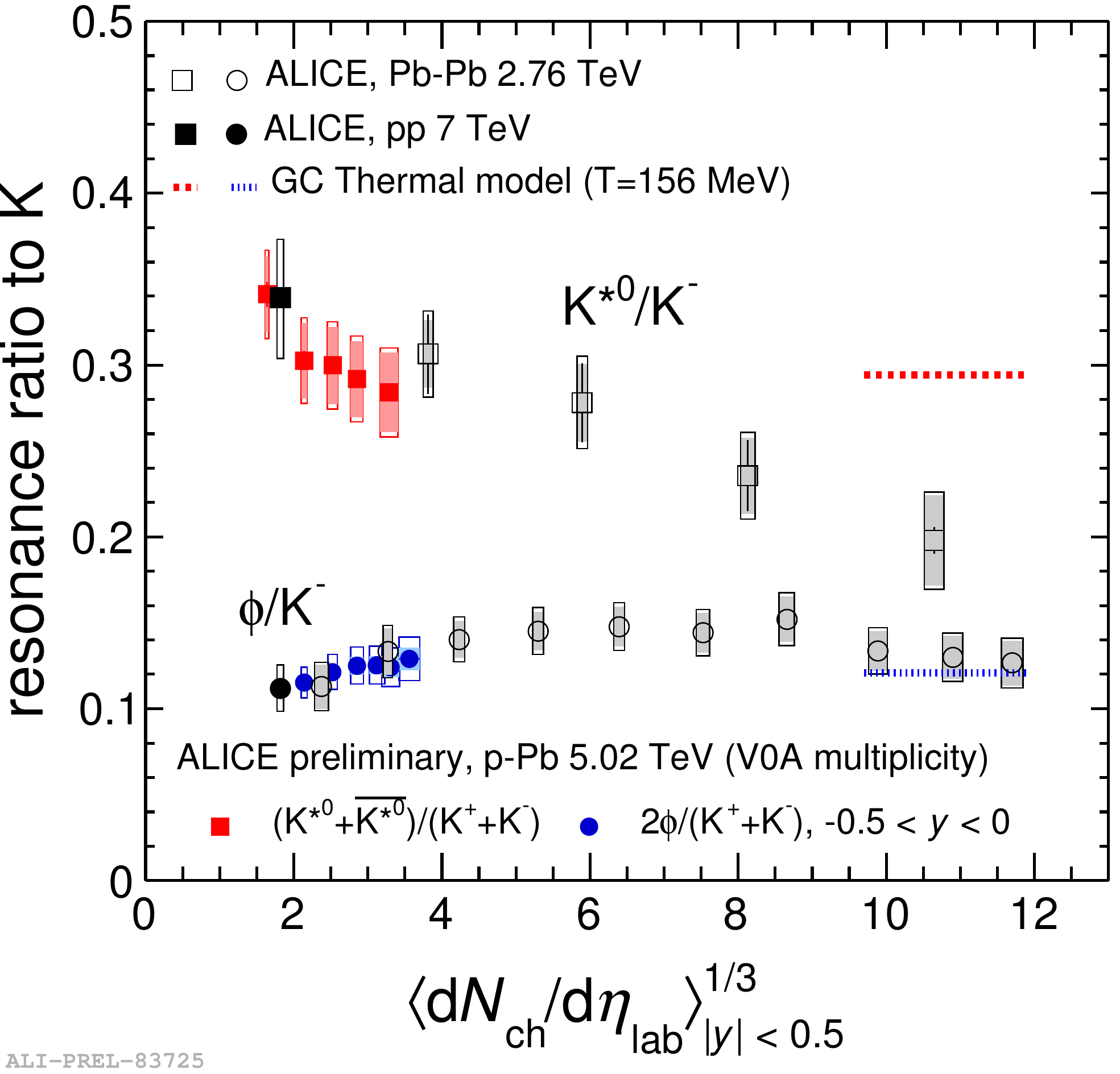}
\caption{ (Left panel) (\p+\ap)/\rmphi~ratio as a function of transverse momentum 
\pT~measured in \pPb~0-5$\%$ (red full circle) and 60-80$\%$ (purple hollow cross) VZERO-A (V0A) multiplicity classes, compared to \pp~(pink hollow circle), 0-10$\%$ (black hollow squares) and 80-90$\%$ (black hollow cross) \PbPb~collisions.
(Right panel)  \simplekstar/\K~and \rmphi/\K~ratios as a function of the cube root of the charged particle multiplicity density \dnchdeta~for \pp, \pPb~and \PbPb~collisions, respectively, at \s= 7~TeV and 
\sNN~=~5.02 and 2.76~TeV. The values given by a grand-canonical thermal model with chemical freeze-out temperature of 156~MeV are also shown~\cite{Stachel}.}
\label{fig:ratios}       % Give a unique label
\end{figure*}

The baryon to meson ratios are useful quantities to study the hadron  production mechanism.  Particularly interesting in this respect is the comparison of the yield of proton and \rmphi, which have a similar mass.  
In the left panel of Fig.~\ref{fig:ratios} the (\p+\ap)/\rmphi~ratio as a function of transverse momentum \pT~for different collisions systems and centrality intervals is shown.  In central (0-10$\%$) \PbPb~collisions this ratio is flat below 3~\gmom, suggesting that the low-\pT~spectral shapes  of the \p~and \rmphi~are mainly determined by the particle mass. The  trend  of distribution of  the (\p+\ap)/\rmphi~ratio  in \pPb~collisions for all event multiplicity classes steeply decreases with \pT, similar to those observed for peripheral \PbPb~collisions and for \pp~collisions. In central \pPb~collisions (i.e. for the 0-5$\%$ V0A multiplicity event class) the ratio shows a hint of flattening for \pT~$<$~1.5~\gmom.

In order to check the presence of a suppression in the production of the resonances and to study whether the strength of the suppression is related to the system size,
the ratios of the \pT-integrated particle yields 
 \simplekstar/\K$^-$ and \rmphi/\K$^-$  have been reported as a function of the cube root of the charged particle multiplicity density $\langle$\dnchdeta $\rangle^{1/3}$, for \pp, \pPb~and \PbPb~collisions, respectively, at \s~=~7~TeV and \sNN~=~5.02 and 2.76~TeV (Fig.~\ref{fig:ratios}, right panel). 
In \pPb~collisions \rmphi/\K$^-$~is  rather independent of the event multiplicity  class and 
\simplekstar/\K$^-$~lies on the interpolation from \pp~to peripheral \PbPb~collisions.
The \rmphi/\K$^-$ ratio in central \PbPb~collisions is almost flat and it is consistent with the estimate of a 
grand-canonical thermal model~\cite{Stachel}, which has a chemical freeze-out temperature of 156~MeV and a baryochemical potential of 0~MeV and does not include 
re-scattering effects. On the contrary, the \simplekstar/\K$^-$~ratio exhibits a clear suppression with the increase of the fireball size, i.e. going from peripheral to most 
central \PbPb~collisions, where the measured ratio is about 60$\%$  of the predicted thermal model value.  
Considering the factor of about 10 between the lifetimes of the two resonances, 
the origin of  the  differences in the \simplekstar~and \rmphi~production could be related to a large modification of the K$^*$ yield due to the pion rescattering mechanism $\sigma$($\pi$,$\pi$),
 which destroys the pion-kaon correlation of the \simplekstar~decay products. 

By assuming  a chemical freeze-out temperature of 156~MeV, a model-dependent estimate of 2~\fmc~as the lower limit of the time between the chemical and kinetic freeze-out has been extracted~\cite{phi_kstar_pbpb} using the measured \simplekstar/\K$^-$ ratio. 

%%%%%%%

\subsection{Transverse momentum spectra and interactions in the hadronic phase}
\label{blast}
\begin{figure*}
\centering
%\begin{figure}[h]
%\begin{minipage}{32pc}
%\vskip +1cm
\includegraphics[width=32pc]{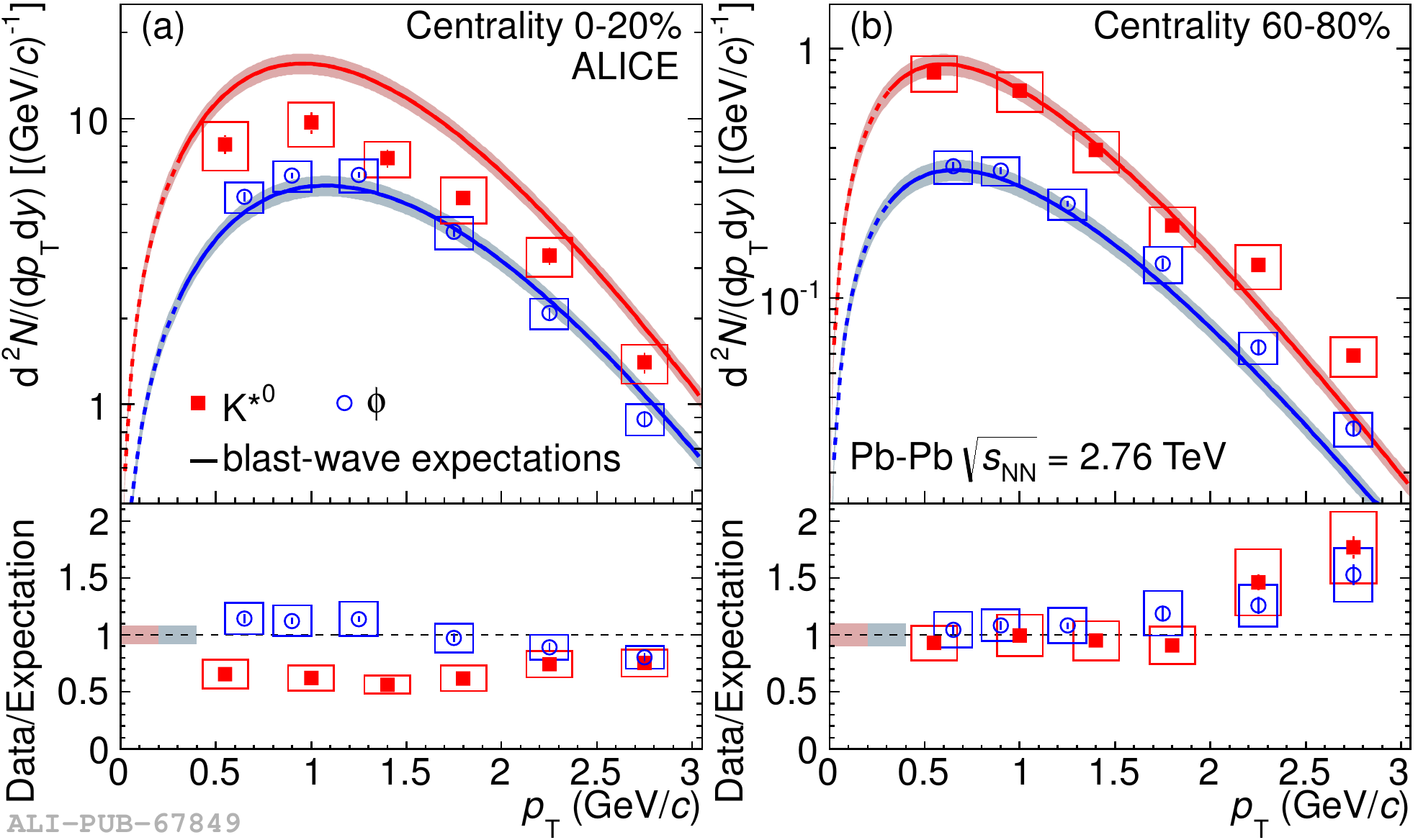}
\caption{\label{fig:blast} Transverse-momentum distribution of \simplekstar~and \rmphi~resonances in central~(a) and peripheral~(b) \PbPb~collisions at \sNN~=~2.76~TeV, compared  to the blast-wave expectation (see text). The lower panels show the ratios of the measured distributions to the prediction.}
\end{figure*}

According to UrQMD calculations~\cite{UrQMD1,UrQMD2} the hadronic rescattering  effect is expected to be momentum dependent with greater strength at low \pT~(\pT~$<$~2~\gmom).
To investigate the \pT~dependence of the observed suppression the blast-wave model~\cite{blast} is used to 
generate an expected transverse-momentum distribution without 
re-scattering effects for \simplekstar~and \rmphi~at kinetic freeze-out. In Fig.~\ref{fig:blast} the transverse momentum distribution of \simplekstar~and \rmphi~resonances in central (0-20$\%$) and peripheral (60-80$\%$)
\PbPb~collisions at \sNN~=~2.76~TeV are compared to the blast-wave prediction for the spectral shape.  The parameters of the blast-wave curves are obtained from a simultaneous fit to the \pT~distributions of charged particles (pions, kaons and protons) in \PbPb~collisions at the same collision energy~\cite{blast_ALICE}. The curves are normalized to the expected resonance yields estimated by multiplying the measured yield of charged kaons in \PbPb~collisions~\cite{blast_ALICE} by the 
\simplekstar/\K~and \rmphi/\K~ratios given by a thermal-model fit to ALICE data~\cite{Stachel}. 
In the \pT~range less than 2~\gmom~the \rmphi~data are satisfactorily described by the prediction in both central and peripheral collisions. The same conclusions hold for \simplekstar~in peripheral collisions, where the data/theory ratio does not appear to deviate significantly from unity.  On the other hand, for 
\pT~$<$~2~\gmom~in central collisions, the \simplekstar~appears suppressed by a factor 0.6.  The deviation from unity is about 3 times larger than the uncertainties, suggesting that \simplekstar~has undergone 
non-negligible re-scattering effects.

%%%%%%%%%%%%%%%%%%%%%%%

%\end{minipage} 
%\end{figure}

\begin{figure}
\centering
\includegraphics[width=8cm]{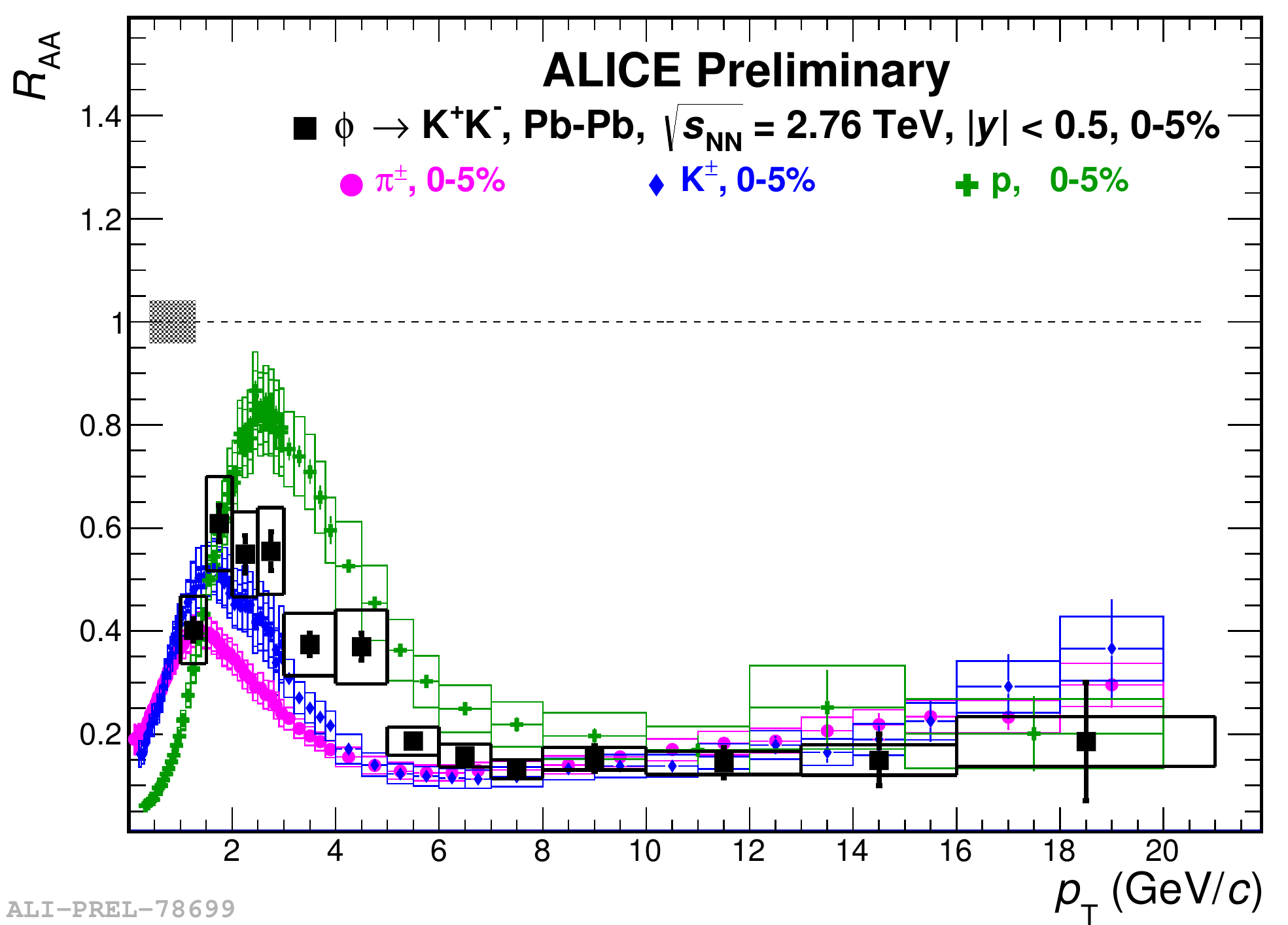}
\caption{ Nuclear modification factor of \rmphi~in 0-5\%  central \PbPb~collisions at \sNN~=~2.76~TeV (\RAA)  compared to that of identified stable hadrons~\cite{RAA2}.}
\label{fig:RAA_phi}       % Give a unique label
\end{figure}
\begin{figure}
\centering
\includegraphics[width=8cm]{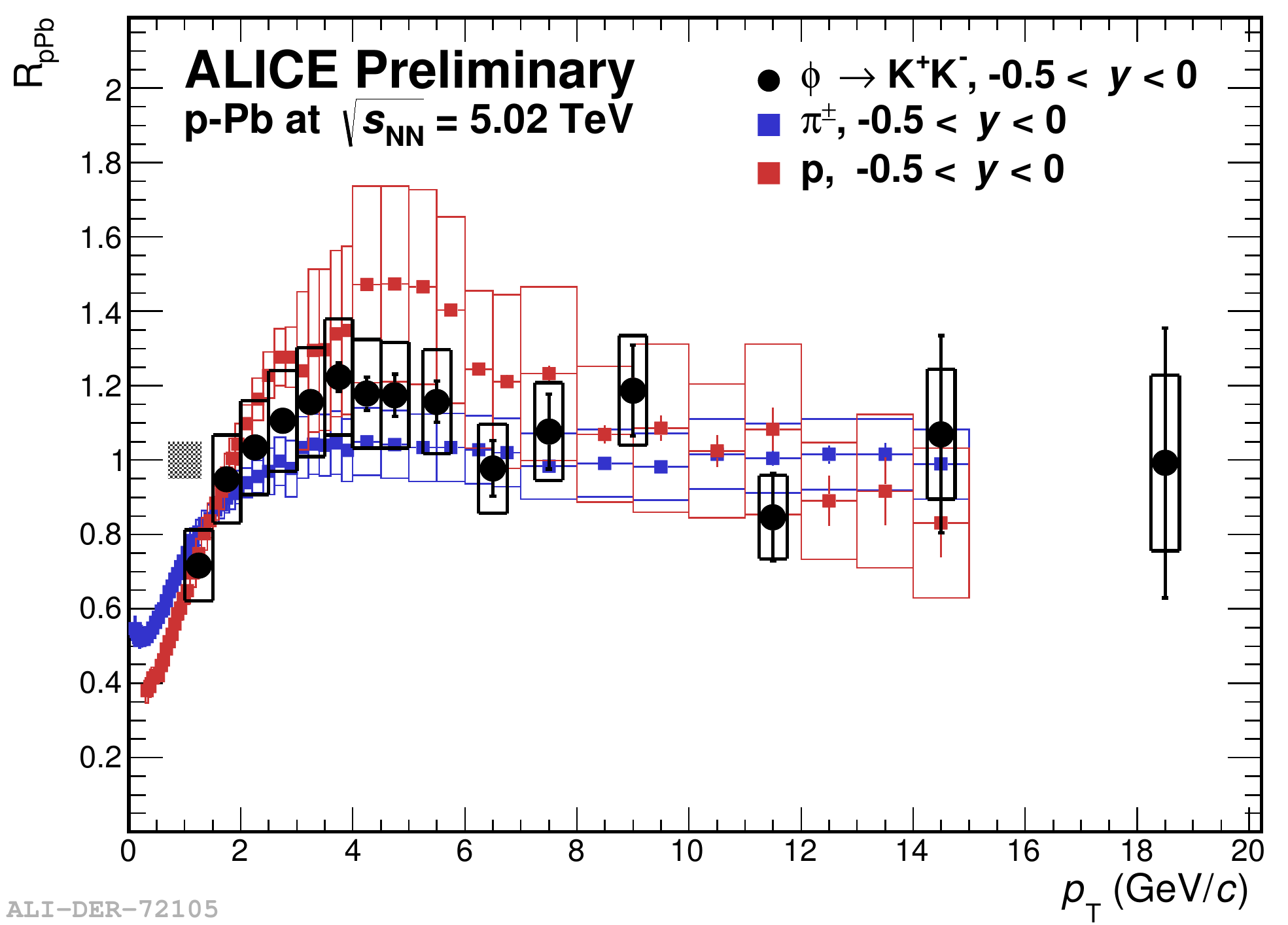}
\caption{ Nuclear modification factor of \rmphi~in minimum bias \pPb~collisions at
\sNN~=~5.02~TeV (\RpPb) compared to that of identified stable hadrons.}
\label{fig:RpPb}       % Give a unique label
\end{figure}
\subsection{Nuclear modification factor}
\label{NMF}
Parton in-medium energy loss  at high \pT~ is usually studied using the so-called nuclear  modification factor \RAA.
It is defined as \RAA (\pT)=$\frac{\mathrm{d}N_{\mathrm{AA}}/\mathrm{d}\pT}{\langle T_{\mathrm{AA}}\rangle \mathrm{d}\sigma_{\mathrm{pp}}/\mathrm{d}\pT}$, 
where $N_{\mathrm {AA}}$ and  $\sigma_{\mathrm {pp}}$ represent the charged particle yield in nucleus-nucleus collisions and the cross section in pp collisions, respectively.  
$T_{\mathrm {AA}}$ is the nuclear overlap function, computed in the framework of a Glauber model~\cite{glauber}.
In central AA collisions a suppression of the production of high \pT~particles has been observed already at RHIC energies.
An increased suppression has been reported by  ALICE~\cite{RAA1,RAA2} at LHC energies, consistent with  the formation of a coloured, dense fireball at these collision energies.

%\begin{figure}
%\centering
%\includegraphics[width=8cm]{Figures/2014-May-17-RAA_KStar_chrg_had_05_v4.pdf}
%\caption{ Nuclear modification factor of \simplekstar~in 0-5\%  central \PbPb~collisions at \sNN~=~2.76~TeV (\RAA)  %compared to that of charged particle.}
%\label{fig:RAA_kstar}       % Give a unique label
%\end{figure}

The nuclear modification factor \rmphi~has been computed for \PbPb~collisions at \sNN~=~2.76~TeV
(\RAA) and for \pPb~collisions at \sNN~=~5.02~TeV (\RpPb). 
In Fig.~\ref{fig:RAA_phi} the  nuclear modification factor of \rmphi~in 0-5\%  central \PbPb~collisions at \sNN~=~2.76~TeV  is shown compared to that of identified stable hadrons.
In central collisions the suppression at  \pT~$>$~8~\gmom~of the \rmphi~is consistent with that measured for the stable hadrons  (\pion, \K~ and \p), thus supporting once more the observation of the flavour-independence of the partonic energy loss in the medium. However, a large baryon/meson dependence is observed for low \pT~in particular \RAA(\p)~$>$~\RAA(\pion). It is worth noting that  the \RAA~of the \rmphi~meson is slightly larger than that of \pion~and lower than the \RAA~of the \p. 
%  In central \PbPb~collisions at low \pT~(\pT~$<$~2~\gmom) the \simplekstar~is more suppressed than the charged %hadrons (Fig.~\ref{fig:RAA_kstar}). This could be an indication of the importance for this particle of the rescattering %effects in the low transverse momentum range.   
  The production of \rmphi~in minimum bias \pPb~and \pp~collisions is compared by computing the \RpPb~(Fig.~\ref{fig:RpPb}).  
  The reference \pp~spectrum at \s~=~5.02~TeV has been obtained from interpolation of the spectra measured in \pp~at 2.76~TeV and 7~TeV, following the same procedure described in~\cite{knichel_QM2014} for identified charged hadrons.  
  The trend of the \RpPb~for the \rmphi~exhibits a moderate Cronin peak (reaching a value of about 1.2) for 3~$<$~\pT~$<$~6~\gmom.  Moreover, no Cronin peak is observed for the pions, while a stronger Cronin peak is present for the protons in the same \pT~range. For the \rmphi~and the stable hadrons no suppression is seen at high-\pT~(\pT~$>$~8~\gmom) in \pPb~collisions compared to \pp.

\section{Conclusions}
\label{conclusion}
The latest results on \kstar~and \phir~resonance production, measured by the ALICE detector in \pPb~and 
\PbPb~collisions at \sNN~=~2.76~TeV and 5.02~TeV, respectively, have been presented. 

 In central \PbPb~collisions the \meanpT~of \simplekstar~and \rmphi~is compatible with that of protons,
while  in \pPb~collisions  the mean \pT~of the resonances does not follow the  same mass ordering as for long lived particles.
% and a the distribution of the \meanpT~as a function of the particle mass suggests the possibility of two %different trends for mesons and baryons.

The ratios of resonances to stable hadrons have been measured and compared in different collision systems. 
The \simplekstar/\K$^-$ ratio is suppressed in central \PbPb~collisions, consistent with substantial re-scattering of \simplekstar~decay daughters in the hadronic phase, while the \rmphi/\K$^-$ is not suppressed consistently with \rmphi~longer lifetime.  The  (\p+\ap)/\rmphi~ratio is flat for  \pT~$<$~3-4~\gmom~in central \PbPb~collisions,  suggesting that the low-\pT~spectral shapes  of the \p~and \rmphi~are mainly determined by the similar particle mass.
In \pp~and in peripheral \PbPb~collisions as well in \pPb~collisions for all event multiplicity classes a steep decrease with \pT~is observed for the (\p+\ap)/\rmphi~ratio.  Only  in central \pPb~collisions (i.e. for 0-5$\%$ VZERO-A (V0A) multiplicity event class) and for low \pT~(\pT~$<$~1.5~\gmom) the ratio shows a hint of flattening, which, as for \PbPb~collisions, could suggest  the onset of a collective behaviour.

The nuclear modification factor of the \rmphi~has been computed for \PbPb~collisions at \sNN~=~2.76~TeV
and for \pPb~collisions at \sNN~=~5.02~TeV. In central \PbPb~collisions, high-\pT~resonances are strongly suppressed as for other stable hadrons, while for the \rmphi~and the stable hadrons no suppression is seen at high-\pT~(\pT~$>$~8~\gmom) in \pPb~collisions compared to \pp.
 
%
% BibTeX or Biber users please use (the style is already called in the class, ensure that the "woc.bst" style is in your local directory)
% \bibliography{name or your bibliography database}

\begin{thebibliography}{}
%
\bibitem{qgp1} S.~Borsanyi  \textit{et al.}, J. High Energy Phys. \textbf{11}, 077 (2010)
\bibitem{qgp1n} S.~Borsanyi \textit{et al.}, J. High Energy Phys. \textbf{09}, 073 (2010) 
\bibitem{qgp2} P.~Petreczky, Proc. of Science (Confinement X) \textbf{028} (2012)
\bibitem{chiral1} G.E.~Brown and M.~Rho, Phys. Rep. \textbf{363}, 85 (2002)
\bibitem{chiral2} R.~Rapp, J.~Wambach and H.~van~Hees, \textit{Relativistic Heavy Ion Physics}, (ed. 
R.~Stock, Spring Berlin Heidelberg, 2010) 134-175
\bibitem{alice} K.~Aamodt  \textit{et al.} (ALICE Collaboration), J. Instrum. \textbf{3}, S08002 (2008). 
\bibitem{phi900} B.~Abelev  \textit{et al.} (ALICE Collaboration), Eur. Phys. J. C \textbf{71}, 1594 (2011)
\bibitem{phi_kstar_pp} B.~Abelev  \textit{et al.} (ALICE Collaboration), Eur. Phys, J. C \textbf{72}, 2183 (2012)
\bibitem{phi_kstar_pbpb} B.~Abelev  \textit{et al.} (ALICE Collaboration), Phys. Rev. C \textbf{91}, 024609 (2015)
\bibitem{Sigmastar_pp} B.~Abelev  \textit{et al.} (ALICE Collaboration), Eur. Phys. J. C \textbf{75}, 1 (2015)
\bibitem{pdg} J.~Beringer \textit{et al.} (Particle Data Group), Phys. Rev. D \textbf{86}, 010001 (2012)
\bibitem{alice_perf} B.~Abelev  \textit{et al.} (ALICE Collaboration), Int. J. Mod. Phys. A  \textbf{29}, 1430044 (2014) 
\bibitem{glauber} M.~Miller \textit{et al.}, Ann. Rev. Nucl. Part. Sci.  \textbf{57}, 205 (2007)
\bibitem{RAA1} K. Aamodt \textit{et al.} (ALICE Collaboration), Phys. Lett. B  \textbf{696}, 30 (2011) 
\bibitem{RAA2} B.~Abelev \textit{et al.} (ALICE Collaboration), Phys. Lett. B  \textbf{720}, 52 (2013)
\bibitem{knichel_QM2014} M.~Knichel (for the ALICE Collaboration),  Nucl. Phys. A  \textbf{931}, 309 (2014)
\bibitem{centrality} B.~Abelev \textit{et al.} (ALICE Collaboration), Phys. Rev. Lett.  \textbf{106}, 032301 (2011)
\bibitem{multpPb} B.~Abelev \textit{et al.} (ALICE Collaboration), Phys. Lett. B \textbf{728}, 25 (2014)
\bibitem{Breit} J.~Adams \textit{et al.} (STAR Collaboration), Phys. Rev. C \textbf{71}, 064902 (2005)
\bibitem{Tsallis} C.~Tsallis, J. Stat. Phys. \textbf{42}, 479 (1988)
\bibitem{Bellini_QM2014} F.~Bellini  (for the ALICE Collaboration), Nucl. Phys. A  \textbf{931}, 846 (2014)
\bibitem{blast_ALICE}  B.~Abelev  \textit{et al.} (ALICE Collaboration), Phys. Rev. C \textbf{88}, 044910 (2013)
\bibitem{radius1} K.~Aamodt \textit{et al.} (ALICE Collaboration), Phys. Lett. B \textbf{696}, 328 (2011)
\bibitem{radius2} M.A.~Lisa \textit{et al.}, Annu. Rev. Nucl. Part. Sci. \textbf{55}, 357 (2005)
\bibitem{meanpT}B.~Abelev \textit{et al.} (ALICE Collaboration), Phys. Lett. B \textbf{727}, 371(2013).
\bibitem{Stachel}  J.~Stachel  \textit{et al.}, J. Phys. Conf. Ser. \textbf{509}, 012019 (2014) 
\bibitem{UrQMD1} S.~Bass \textit{et al.}, Prog. Part. Nucl. Phys. \textbf{41}, 255 (1998)
\bibitem{UrQMD2} M.~Bleicher \textit{et al.}, J. Phys. G \textbf{25}, 1859 (1999)
\bibitem{blast} E.~Schnedermann, J.~Sollfrank and U.~Heinz, Phys. Rev. C \textbf{48}, 2462 (1993).
\end{thebibliography}

\end{document}